# Enhancement of magnetoresistance in manganite multilayers

A.Venimadhav[#], M.S Hegde[#], V. Prasad[*] and S.V. Subramanyam[*]

#Solid state and structural chemistry unit, Indian institute of science, Bangalore-12, India
*Department of Physics, Indian institute of science, Bangalore-12, India

---

mshegde@sscu.iisc.ernet.in

madhav@sscu.iisc.ernet.in

# Abstract

Magnanite multilayers have been fabricated using $La_{0.67}Ca_{0.33}MnO_3$ as the ferromagnetic layer and $Pr_{0.7}Ca_{0.3}MnO_3$ and $Nd_{0.5}Ca_{0.5}MnO_3$ as the spacer layers. All the multilayers were grown on $LaAlO_3$ (100) by pulse laser deposition. An enhanced magnetoresistnace (defined $(R_H - R_0)/R_0$ ) of more than 98% is observed in these multilayers. Also a low field magnetoresistance of 41% at 5000 Oe is observed in these multilayer films. The enhanced MR is attributed to the induced double exchange in the spacer layer, which is giving rise to more number of conducting carriers. This is compared by replacing the spacer layer with $LaMnO_3$ where Mn exists only in 3+ state and no enhancement is observed in the $La_{0.67}Ca_{0.33}MnO_3$ / $LaMnO_3$ multilayers as double exchange mechanism can not be induced by external magnetic fields.

# 1 Introduction

Manganite perovskite oxides has been the subject of interest after the discovery of colossal magnetoresistance[1,2]. These manganites with the generic formulation $Ln_{1-x}A_xMnO_3$ (for $0.2< x<0.5$) shows high negative magnetoresistance at curie temperature. There have been extensive experimental and theoretical studies to understand the origin of the anomalous magnetotransport phenomenon. Qualitative explanation is given by double exchange mechanism. Another intriguing phase, the charge ordered state has been found to exist in insulating $Ln_{1/2}A_{1/2}MnO_3$. The charge ordered state is characterized by the real space ordering of $Mn^{3+}/Mn^{4+}$ in the mixed valent manganite. The charge ordered state can be melted to ferromagnetic metallic (FMM) state by applying external magnetic field [3]. The melting of the charge ordered state is associated with a huge magnetoresistance of thousand folds.

Artificial superlattice is another interesting topic in the field of ultra thin films. Giant magnetoresistance has been observed in the ferromagnetic metal and non ferromagnetic metal superlattices [4]. Manganite superlattices have been studied by many researchers in the last few years. The magnetic transport property of the ferromagnetic oxides is observed to vary according to the magnetic/non magnetic spacer layers materials such as $SrTiO_3$, $SrRuO_3$, $La_{1-x}MnO_{3+\delta}$, $Gd_{0.7}Ca_{0.3}MnO_3$, $LaNiO_3$, $La_{0.55}Sr_{0.45}MnO_3$, $La_{1/3}Ca_{2/3}MnO_3$ and $La_{0.6}Sr_{0.4}FeO_3$, [5-12]. The multilayer study of manganites is important as it has a direct application in the recording media technology. The ultimate goal to be achieved for technological interest is a high magnetoresistance in low magnetic fields at room temperature. Interestingly manganite oxide multilayers have shown enhanced magnetoresistance effect [5-8]. For instance in $La_{0.67}Ca_{0.33}MnO_3/SrTiO_3$ multilayers where $SrTiO_3$ is a non magnetic insulating oxide exhibits a large MR of more than 85% at low temperatures below 100 K[5]. In $La_{0.67}Ca_{0.33}MnO_3/SrRuO_3$ superlattices a substantial enhancement of magnetoresistance is observed at low temperatures. This is explained by the induced magnetic non uniformity near the interfaces due to disorder, which serve as additional spin dependent scattering centers[6].

Below the transition temperature a large internal field is developed in the CMR materials such as $La_{0.67}Ca_{0.33}MnO_3$ (LCMO). Making a multilayer of LCMO and charge ordered material, it may be possible to melt the charge ordered state by a small magnetic field below the transition temperature of LCMO. Due to this the complete multilayer become a ferromagnetic metal and can show a huge manetoresistance. In this paper we report the magnetotransport properties of the multilayers made with ferromagnetic and the charge ordered manganites for the first time. We have fabricated LCMO/$Pr_{0.7}Ca_{0.3}MnO_3$, LCMO/$Nd_{0.5}Ca_{0.5}MnO_3$ multilayers. Indeed we have obtained magnetoresistance of 98% in these multilayers. Also a magnetoresistance of 41% is seen at 0.5 T. These results are compared with multilayer made with a non charge ordered insulator material $LaMnO_3$, which has not shown any enhancement in MR.

## 2 Experiment

Stoichiometric targets of $La_{0.67}Ca_{0.33}MnO_3$ (LCMO), $Pr_{0.7}Ca_{0.3}MnO_3$ (PCMO), $Nd_{0.5}Ca_{0.5}MnO_3$ (NCMO) and $LaMnO_3$ (LMO) were prepared by solid state route and characterized by X-ray diffraction. The polycrystalline targets were used for the pulse laser deposition (PLD). The PLD system is equipped with a multi target holder for the insitu superlattice deposition. All the films were grown on $LaAlO_3$ (100) substrate. The single layers and multilayers were deposited under similar experimental condition. Substrate temperature was maintained at 750-780 $^0$C and the oxygen pressure in the chamber was kept at 330 mTorr. All the films were annealed insitu at 750 $^0$C in 500 Torr of oxygen for one hour. We have made single layer films of LCMO, PCMO and NCMO. In the single layer grown films the thickness of LCMO was 1000 Å and the other films were 3000 Å thick. Multilayers have been fabricated using LCMO as the ferromagnetic layer by varying the spacer layers with PCMO and NCMO. There are 20 bilyaers in each multilayer. To compare these multilayers we have made another set of multilayers with LCMO and LMO. In all the multilayers thickness of the LCMO was kept constant at 50 Å. LCMO/PCMO and LCMO/LMO superlattices were grown with different thickness of PCMO (10 Å,15 Å and 20 Å) and LMO (10 Å and 15 Å) respectively. In LCMO/NCMO multilayer NCMO spacer layer thickness was 10 Å. The structure of the films were characterized by Seimens X ray diffractometer. Resistivity measurements were performed in standard four probe configuration. Magnetic property of the multilayers has been studied by faraday force balance. Magnetoresistance measurements were carried out up to a magnetic field of 8T.

## 3 Results and discussion

The LCMO film on LAO (100) showed perovskite type cubic structure with a lattice parameter of 3.9 Å. The out of plane parameter calculated for PCMO and NCMO is 3.85 Å. The growth of PCMO and NCMO is along (101) direction on LAO[13,14]. Super lattice peeks are observed for the multilayers. Also the 0-360 phi scan performed on the multilayers proves the epitaxial growth. Single layer electrical property of PCMO and NCMO showed insulating behavior. Resistivity Vs temperature plots of single layer PCMO and NCMO films with and without magnetic field are shown in Fig 1. Though metallic behavior was not observed in presence of 8T magnetic field, yet they show a high negetive magnetoresistance to the applied field. Charge ordering behavior was not observed in any of the films. Single layer LCMO film showed insulator – metal transition ($T_{IM}$) at 250 K as given in the inset of Fig.1 which agrees well with the epitaxial films reported in the literature[14]. LMO film showed insulating behavior.

The resistance Vs temperature (R Vs T) plot of the multilayers LCMO/PCMO, LCMO/NCMO and LCMO/LMO are shown in Fig. 2. Though the multilayers showed $T_{IM}$, the transition was shifting to the lower temperature side of the LCMO transition. The $T_{IM}$ of LCMO/NCMO observed 160 K is lower than the LCMO/PCMO multilayer at 225K. Inset of Fig. 2 shows the thickness dependence of $T_{IM}$ in PCMO multilayers where $T_{IM}$ decreased with increase in the thickness of PCMO layer. LCMO/LMO multilayer has also shown a similar R vs T behavior. It is evident from the results that, the insulator to metal (IM) transition temperature of the multilayers decreases either by bringing a high resistive material between the LCMO layers or by increaseing the thickness of the high resistive material (PCMO, NCMO as well as LMO).

Susceptibility of the multilayer samples has been measured in a magnetic field of 500 Oe. Susceptibility vs temperature plots of the multilayers are shown in Fig 3. LCMO/PCMO multilayer has shown a curie transition 225K and saturates below 150 K. Though there was a ferromagnetic transition in LCMO/NCMO multilayer, magnetisation is not saturated till the measured lowest temperature. The $T_c$ of PCMO and NCMO multilayers are almost the same. This implies that the ferromagnetic contribution is mainly arising from the LCMO layer. In case of LCMO/LMO though magnetization start raising up from 230 K, a sharp transition occurs at 150 K.

We have measured the magnetoresistance of the samples up to a magnetic field of 8T. Fig 4. Shows a typical magnetoresistance plot of the LCMO/PCMO multilayers measured at different temperatures. The behavior is quite similar to the single layer material in which, a $H^2$ dependance of magnetoresistance is normally seen above the $T_{IM}$ and an exponential kind of behavior below the transition. The magnetoresistance reduces on moving from either side of the transition temperature. Fig 5 shows the comparison of maximum magnetoresistance observed in the multilayer films near their respective $T_{IM.}$ LCMO film has showed a maximum magnetoresistance of 85% at 250K. The LCMO/PCMO and LCMO/NSMO multilayers have shown more than 98% magnetoresistance just below their transition temperature. Note from the fig. 5 that at 0.5T, MR of ~ 41% is seen in these multilayers. A comparison of the curvatures of the single LCMO and the multilayers clearly shows the enhancement of magnetoresistance in the LCMO/PCMO and LCMO/NCMO multilayers. We have not observed any hysteresis in the plot of magnetoresistance vs magnetic field. Also magnetoresistance measurement were done on LCMOP/PCMO multilayers by applying the magnetic field in both parallel and perpendicular directions to the substrate. No anisotropy is observed in these measurements. In the same plot magnetoresistance vs H behavior of LCMO/LMO multilayer is shown with two different LMO thickness. The magnetoresistance of these multilayers was small comparing to single LCMO film. Increasing the LMO thickness further reduces the magnetoresistance to 70%. There is no considerable low field magnetoresistance observed in LCMO/LMO multilayers.

The significant observation in this study is the enhancement of magnetoresistance up to 98% in the LCMO/PCMO and LCMO/NCMO multilayers. The other observations to be understood are the decrease of $T_{IM}$ and $T_C$ in the multilayers compared to the LCMO film. For the current in plane (CIP) measurements the magnetotransport depends on the mean free path of the conducting electrons in the magnetic layer (LCMO) and the thickness of the spacer layer (PCMO). One can expect more conductivity when the current samples through more number of magnetic layers. In other words the mean free path is larger than the spacer layer thickness. One could easily expect this to happen as the mean free path of the manganites is small, of the order of few unit cells. Hence increasing PCMO thickness will reduce the $T_{IM}$ of the multilayers. Also incresing the resistivity of the spacer layer decreases the $T_{IM}$. This seems to be the reason why $T_{IM}$ of LCMO/NCMO is less than LCMO/PCMO mutilayer. The interfacial strain effects may also bring down the $T_{IM}$ and $T_C$ in the multilayer structure [6]. In a multilayer, by introducing a non magnetic spacer layer between the ferromagnetic layers alters the orientation of the moments in the ferromagnetic layers and there by changes the strength of the ferromagnetic coupling. This can also bring down the $T_C$ in the multilayers.

In general, decease in $T_{IM}$ enhances the magnetoresistance in CMR materials [15-17]. All the multilayers showed a decrease in $T_{IM}$. While LCMO/PCMO and LCMO/NCMO showed enhancement in magnetoresistance and no enhancement observed in LCMO/LMO multilayers. This signifies the role of the spacer layer in the enhancement of magnetoresistance. The present study should be contrasted with the previous multilayer studies made with $SrTiO_3$ [5] and $SrRuO_3$ [6] as the spacer layers where an enhanced MR was observed at low temperatures well below $T_{IM}$. This enhancement was attributed to the spin dependent scattering across the disordered boundary. This reason can be ruled out in the present case as a clean interface is expected in the multilayer structure as both the layers are manganites. The multilayers made with manganites as the spacer layer may behave differently. Hence the physical origin of the magnetoresistance observed here is distinct from such studies reported earlier [6-8]. We have examined the effect of the nature of the spacer layer on magnetoresistance in multilayers by comparing the above three spacer layers (PCMO, NCMO and LMO) and a possible explanation is given below.

It is well known that the manganese ions are playing the main role in the transport properties of the manganites. PCMO and NCMO show charge ordering in bulk solids. The charge ordering was not seen in the thin films of these samples [13]. Though PCMO (NCMO) does not show charge ordering, the charge carriers in the $Mn^{3+}$ and $Mn^{4+}$ states are localized in the insulating state and shows negative magnetoresistance in response to the applied external magnetic field. LCMO layers are ferromagnetic metals below their transition temperature (250K) and the large internal field of LCMO layers acts on the spacerlayer. Application of an external magnetic field further helps in the alignment of the LCMO layers and acts together on spacer layer. The added field effect may order $Mn^{3+}$ and $Mn^{4+}$ ferromagnetically in the spacer layer leading to $Mn^{3+}$-O-$Mn^{4+}$ double exchange. This reduces the resistance of the multilayer giving a huge magnetoresistance. On the other hand the multilayer with LMO as the spacer layer has not shown any enhancement in MR. Though LMO is an insulator it contains manganese only in 3+ state and any external magnetic field can not drive double exchange in LMO. Therefore no enhancement in the MR is expected in this multilayer. Further, increase in the thickness of the LMO layer reduces the magnetoresistance as clearly seen from Fig 5. Hence spacer layer material should be chosen in such a way that it releases more number of conducting carriers by the application of the magnetic field to obtain large magnetoresistance. In the resent study on $La_{0.7}Ca_{0.3}MnO_3$ / $Gd_{0.7}Ca_{0.3}MnO_3$ multilayer[8] the observed low temperature low field magnetoresistance has been attributed to the interfacial strain produced in the layered structure. In the present study also the possibility of interfacial strain may act on the LCMO layer by the spacer layer. But this can not explain why LCMO/LMO multilayer did not show any enhancement. Also in the present study the low field magnetoresistance was observed only at $T_{IM}$. No low field magnetoresistance is observed at low temperatures. Hence the MR effect may seem to depend on the electronic property of the spacer layer. This study gives a new direction to obtain high magnetoresistance in manganite multilayers.

In conclusion we have made multilayers of LCMO/PCMO and LCMO/NCMO and studied their electrical transport property in presence of magnetic field. All the samples showed insulator to metal transition. Magnetoresistance of more then 98% is observed in these films. On the other hand no enhancement is observed in multilayers replacing LMO as the spacer layer. The reason for the enhanced MR suggested is due to the induced double exchange mechanism in PCMO (NCMO) by applying the magnetic field.

We acknowledge Department of Science and technology, India for the financial support.

Fig.1 Resistivity versus temperature plot of PCMO and NCMO films with and without magnetic field. Inset shows the ρ vs T behavior of LCMO film with and without magnetic field.

Fig. 2 R vs T plot of the multilayers (□)LCMO$_{50}$/NCMO$_{10}$ (○)LCMO$_{50}$/PCMO$_{10}$ (△) LCMO$_{50}$/LMO$_{10}$. Inset shows the R vs T plot of LCMO/PCMO multilayers with different PCMO thickness (▽) LCMO$_{50}$/PCMO$_{10}$ (◇) LCMO$_{50}$/PCMO$_{15}$ (+) LCMO$_{50}$/PCMO$_{20}$

Fig. 3 Normalised magnetization versus temperature is shown for the multilayers.

Fig. 4 Typical magnetoresistance Vs applied magnetic plot for LCMO$_{50}$/PCMO$_{10}$ multilayer at different temperatures.

Fig. 5 Magnetoresistance as a function of the applied field is ploted for the multilayers. The curves represent the maximum magnetoresistance obtained in the multilayers( for all the samples its is near their T$_{IM}$).

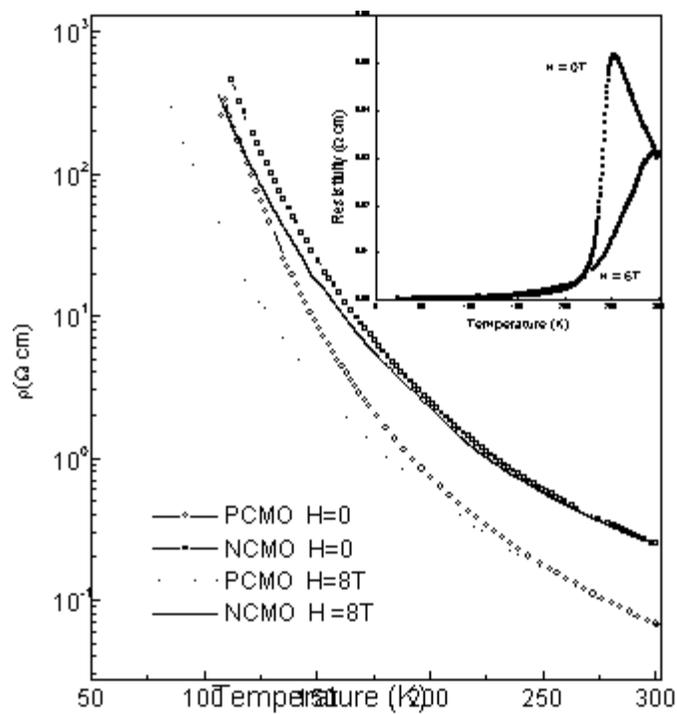

Fig.1
Venimadhav et al,.

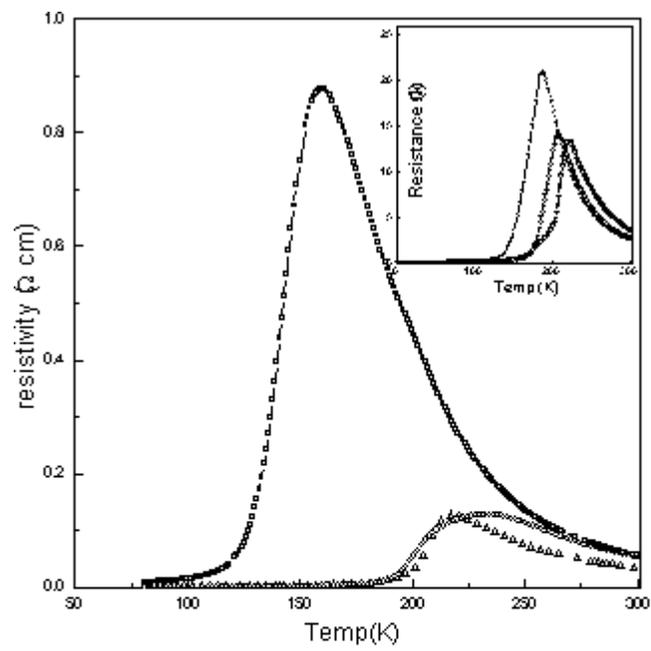

Fig2
Venimadhav et al,.

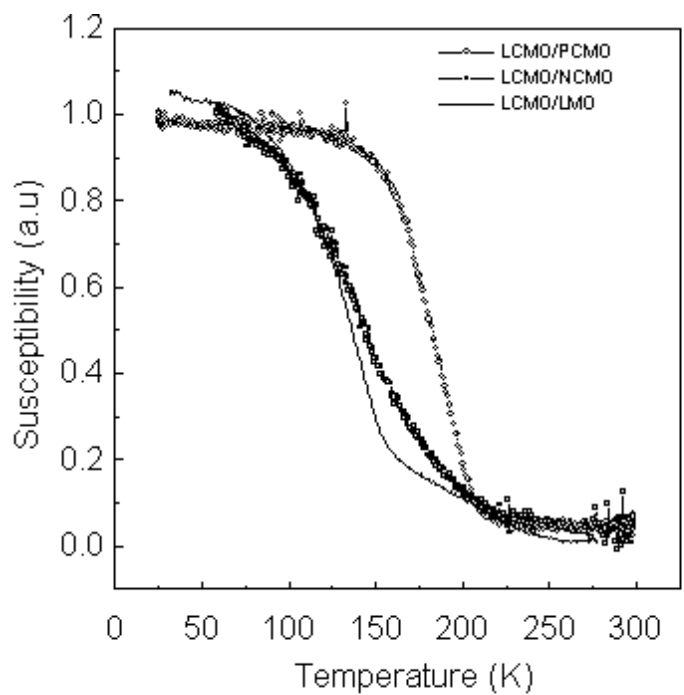

Fig. 3
Venimadhav et al,.

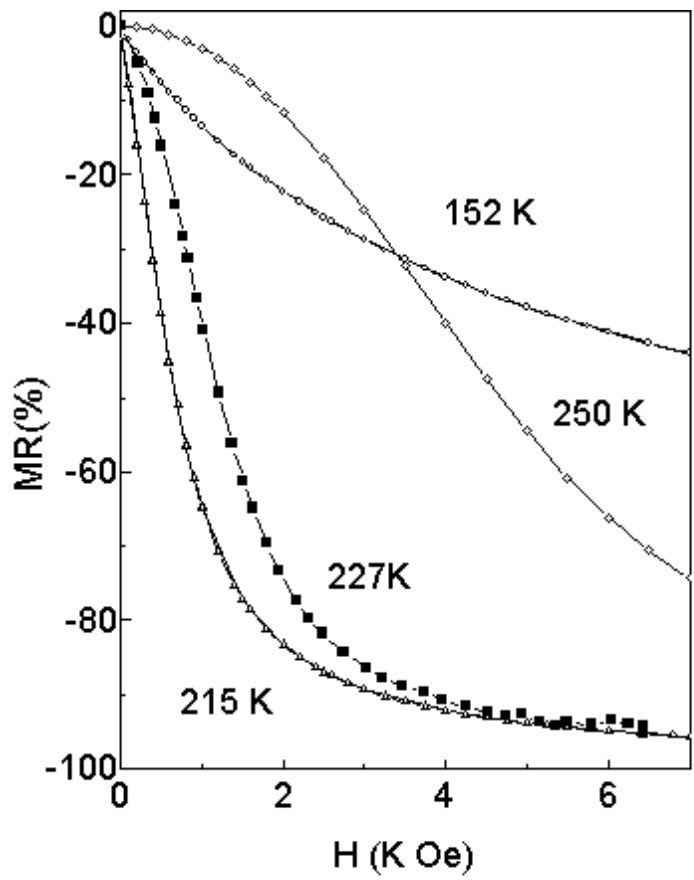

Fig. 4
Venimadhav et al,.

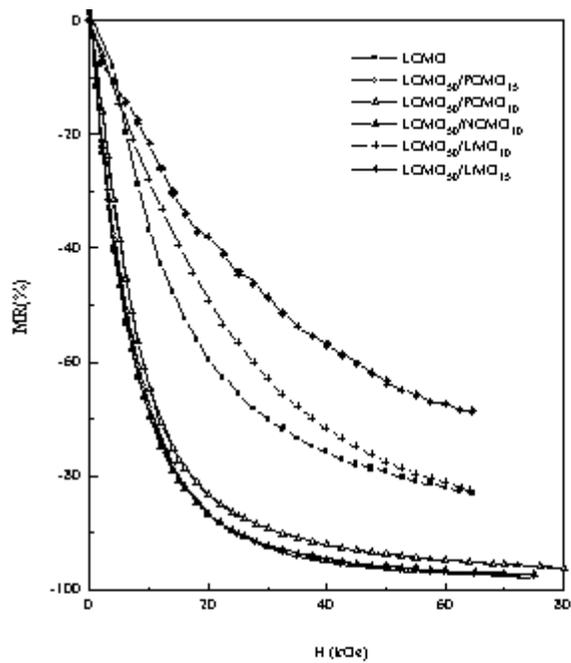

Fig.5
Venimadhav et al,.